\documentclass[onecollarge]{svjour2}       
\smartqed  
\usepackage{graphicx}
\usepackage{amsmath}
\usepackage{amsfonts}
\usepackage{color}
\usepackage{bm}
\begin{document}

\title{Homogeneous and Isotropic Turbulence: a short survey on recent developments.\footnote{postprint version, accepted for publication on J. Stat. Phys. DOI 10.1007/s10955-015-1323-9}}


\author{Roberto Benzi      \and
        Luca Biferale 
}


\institute{R. Benzi \at
              Dept. of  Physics and INFN, University of Rome `Tor Vergata', Rome, Italy \\
              \email{benzi@roma2.infn.it}           
           \and
           L. Biferale \at
              Dept. of  Physics and INFN, University of Rome `Tor Vergata', Rome, Italy \\
              \email{biferale@roma2.infn.it}           
}

\date{Received: 29 June 2015 / Accepted: 6 July 2015 }

\maketitle

\begin{abstract}
We present a detailed review of some of the most recent developments on Eulerian and Lagrangian turbulence in homogeneous and isotropic statistics. 
In particular, we review phenomenological and numerical results concerning the issue of universality with respect to the large scale forcing and the
 viscous dissipative physics. We discuss the state-of-the-art of numerical versus experimental comparisons and we discuss the dicotomy between phenomenology based on coherent structures or
on  statistical approaches. A detailed discussion of finite Reynolds effects is also presented. 
\end{abstract}

\section{Introduction}
In this paper we critically review the most relevant physical features of fully developed three dimensional turbulence. We consider the case of
incompressible homogeneous and isotropic turbulence.  
Although this case may be considered too limited,  most of our arguments can also be extended 
in the case of non isotropic turbulence \cite{Hinze:1959iz,Monim:bHWl3078,Biferale:2005cw,Frisch:1995ti}. 

\bigskip

The word turbulence refers to the chaotic behavior, in space and time, of fluid flow. It has been a major breakthrough 
to understand how the complexity of turbulent flows can be described by the Navier Stokes equations:
\begin{equation}
\partial_t {\bf v} + {\bf v} \cdot \nabla {\bf v} = - \frac{1}{\rho} \nabla {\bf v} + \nu \Delta {\bf v}
\label{NS}
\end{equation}
where ${\bf v}$ is the velocity field , $p$ the pressure, $\nu$ the kinematic viscosity and
 $div {\bf v}=0$ for incompressible flows. Nowadays, eq.s (\ref{NS}) can be numerically simulated with high accuracy  and 
we reproduce very accurately  in a computer  what one can measure in a laboratory experiments \cite{Arneodo:1996hf,Arneodo:2008gq}. Turbulence was the
 first and more challenging physical phenomena investigated by numerical simulations. Following Kadanoff, it is fair to say that physics is no longer divided in experimental and theoretical physics: numerical simulations or numerical experiments are new ways to understand complex phenomena. 

\bigskip

Given any external force acting in the system, we can define the rate of energy input and energy dissipation in a turbulent flow. Hereafter we shall denote by $v_0$ the characteristic
scale of velocity fluctuations  and by $L$ the scale characterizing energy input. We can usually describe how much turbulent a flow is, by introducing the
Reynolds number $Re \equiv v_0L/\nu$.  Numerical simulations can be performed for $Re$ as large as $10^{6}$ \cite{doi:10.1146/annurev.fluid.010908.165203} which is also one of the largest Reynolds number for laboratory experiments.  In the next few years, we expect to reach $Re=10^{7}$ for the numerical simulations. So far, there exists no indications that eq.s (\ref{NS}) are wrong, i.e. that there are physical features not described by ({\ref{NS}) for non-relativistic, incompressible and ideal fluids. This is a highly non trivial statement which is based from a detailed and careful comparison between numerical simulations and experimental data and more than 30 years of hard work.

\bigskip

\section{The Kolmogorov theory}

Being a chaotic system, turbulence must be described in a statistical way, i.e. we need to introduce correlation functions and a probability measure for the velocity field. However, turbulence is a system strongly out of equilibrium and the standard tools and ideas of statistical physics are useless: we need to build up a new theoretical framework. Physically, the basic and most fundamental feature of three dimensional turbulent flows is that energy dissipation is independent on the Reynolds number. More precisely, energy dissipation $\epsilon$ for eq.s (\ref{NS}) is given by $\langle \nu (\nabla {\bf v})^{2} \rangle $, where $\langle \dots \rangle$ means an  average in space and time. Since $Re \rightarrow \infty$ is equivalent to $\nu \rightarrow 0$, we expect that 
$
\epsilon \rightarrow 0$ {\it iff}  the velocity gradients are bounded. Experimentally as well as numerically we have that 
\begin{equation}
\label{eq:anomaly}
\epsilon \sim const. \qquad Re \rightarrow \infty
\end{equation}
 which is the so-called {\it dissipative anomaly}, implying that $ \nabla {\bf v}$ grows to infinity. Kolmogorov was the first who clearly highlighted this basic and fundamental feature of turbulence. In his celebrated 1941 paper, Kolmogorov was able to show that for homogenous and isotropic turbulence, the statistical properties of turbulent fluctuations must satisfy the equation \cite{Frisch:1995ti}:
\begin{equation}
\langle \delta v(r) ^ 3 \rangle = - \frac{4}{5} \epsilon r + 6 \nu \frac{d}{dr} \langle \delta v(r) ^2 \rangle
\label{K45}
\end{equation}
 where $ \delta v(r) \equiv ( {\bf v}({\bf x} + {\bf r})-{\bf v}({\bf x})) \cdot {\bf r}/r $ is the longitudinal difference of the velocity field between two points at distance $r$. Eq. (\ref{K45}) is exact and must be  well verified both by numerical simulations and experimental data if stationarity is assumed and (\ref{eq:anomaly}) holds.  In the limit $ \nu \rightarrow 0$, Eq. (\ref{K45}) predicts the existence of two different {\it range} of scales $r$. Introducing the scale $\eta = (\nu^3/\epsilon)^{1/4}$, we can say that for $r \gg \eta$ the velocity fluctuations are controlled by the energy dissipation $\epsilon$ and the scale $r$, while for $r \sim \eta$ dissipation effects become important. The first range of scale ($r \gg \eta$) is called inertial range and the second range is called the dissipation range. 
 
 \bigskip
 Eq. (\ref{K45}) suggests that the probability distribution of turbulent fluctuations at scale $r$ depends only on $\epsilon$ and $r$. Using dimensional arguments one can conclude  that 
 \begin{equation}
 \label{SP}
S_p(r) =  \langle \delta v(r)^p \rangle \sim \epsilon^{p/3} r^{p/3}
 \end{equation}
There are two major points in the Kolmogorov theory. First, eq. (\ref{K45}) is the fundamental prediction of the theory assuming that turbulent fluctuations are statistically isotropic, isotropic  and $\epsilon$ is constant for $Re \rightarrow \infty$. Secondly, eq. (\ref{SP}) can be considered a {\it conjecture} of the theory assuming that the statistical properties of turbulent flows are {\it scale invariant} in the inertial range, where the notion of scale invariant should be interpreted in the same way introduced in the theory of critical phenomena. Note that eq. (\ref{K45}) is true even if scale invariance does not hold. 

\bigskip

The simplest way to check (\ref{SP}) is to compute the quantities 
$$\Gamma_p(r) \equiv S_p(r)/S_2(r)^{p/2}$$ usually referred to as generalized kurtosis. Eq. (\ref{SP}) predicts that $\Gamma_p(r) \sim const$ in the inertial range. This is definitively not observe both in numerical simulations and laboratory experiments where $\Gamma_p(r)$  increase for  $r \rightarrow \eta$, 
see figures (\ref{fig1}) and (\ref{fig2}). This phenomenon is called intermittency. 

\bigskip

In order to understand intermittency in a physical way, it is possible to look at turbulent flows also from the Lagrangian point of view (opposed to the Eulerian one, based on measurements in a fixed reference frame in the laboratory). 
We consider a particle (point like) which is advected by the velocity field ${\bf v}$ and whose trajectory is described by the position ${\bf x}(t)$. We can compute
the lagrangian velocity difference on time $\tau$ defined as $ \delta v (\tau) = | {\bf v}(t+\tau)-{\bf v}(t)|$ and the quantities $S^{L}_p \equiv \langle \delta v(\tau)^p \rangle$, where the superscript $L$ denotes the lagrangian frame. The analogous of the scale dissipation $\eta$ is now the Kolomogorv time scale $\tau_d \equiv (\nu/\epsilon)^{1/2}$.
Dimensional arguments and scale invariance implies that $S^L_p (\tau) \sim (\epsilon \tau)^{p/2}$. Numerical and experimental data shows clearly that $\Gamma^L_p(\tau) \equiv S^{L}_p/(S^L_{2})^{p/2}$ increases when $\tau \rightarrow \tau_d$. Usually, the value of $\Gamma^L_p (\tau_d)$ is much larger than the corresponding Eulerian quantity $\Gamma_p(\eta)$, i.e. we observe much stronger intermittency in lagrangian framework with respect to the eulerian framework, see fig. (\ref{fig1}).  We can also compute the acceleration $a \equiv | d^2{\bf x}/{dt^2}|$. According to scale invariance in the form (\ref{SP}), we can estimate $a \sim \delta v(\tau_d)/\tau_d$ and compare it against data. It turns out that $a$ is one of the most intermittent quantities observed
in turbulent flows: in fig. (\ref{fig4}) we show that  the probability distribution of $a$ develops extremely long tail up to $80$ times the variance! Such a spectacular behavior has a well defined physical interpretation:
from time to time the particle enters a region of extremely large vorticity ${\bf \omega} \equiv rot {\bf v}$, see figure (\ref{fig3}). These regions of large vorticity usually take the form of filaments which extend in space for scales well within the inertial range  and show a cross section of order $10 \eta$. A vortex filament is neither a stationary nor a stable structure: when a particle enters a vortex filament  the value of $a$ is of order $a \sim \omega^2 r_f$ where $\omega$ is the vorticity of the filament and $r_f$ is the cross section. For large $Re$, the vorticity scales as $Re^{1/2}$ and $a$ can easily becomes much larger than its characteristic value outside  filaments. In summary, intermittency in the lagrangian acceleration is related to the existence of vortex filaments and viceversa. 

\bigskip

Vortex filaments have been observed in all turbulent flows. In wall bounded turbulence, vortex filaments are responsible for the drag effect near the wall and control the rate of energy production and dissipation \cite{pope2000turbulence}. The peculiar property of vortex filaments is that they do not carry most of the energy fluctuations in a turbulence flow but they organize the flow around them and the region of the energy dissipation. The existence of vortex filaments open up a completely different scenario for turbulent flows: although Kolmogorov theory (i.e. eq. (\ref{K45})) is correct and in agreement with all existing data, scale invariance as described by eq. (\ref{SP}), or its equivalent form for lagrangian dynamics,  does not properly take into account the complex non linear intermittent dynamics   in the statistical properties of turbulence. Moreover, since vortex filaments extend well within the inertial range, their statistical properties may be related to the large scale forcing. This implies that there may not be a universal way to describe the statistics  of turbulence {\it independent} of the forcing mechanisms. Also, since the cross section of a vortex filaments is not much larger than the dissipation scale, the statistical properties of turbulence may strongly depend on the dissipation mechanism acting at very small scales. In other words, turbulence may not be universal with respect to large scale forcing and small scale dissipation. This is the crucial problem we need to understand in the following.

\begin{figure}[h]
 \begin{center}
 \includegraphics[width=0.90\textwidth]{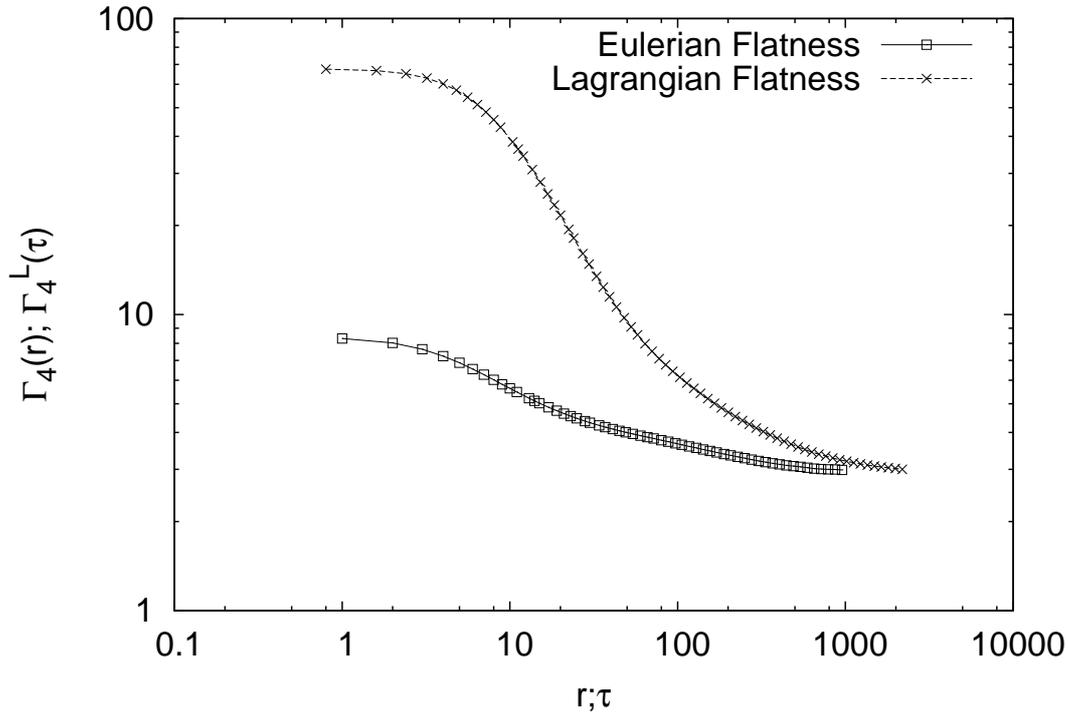}
 \end{center}
   \caption{Flatness for  Eulerian and Lagrangian measurements, 
data are taken from a DNS at $2048^3$ resolution \cite{Bec:2010hj}}
  \label{fig1}
\end{figure}

\bigskip

\section{The multifractal approach}

\bigskip

Even if vortex filaments are crucial in understanding intermittency, there is no reason to assume {\it a priori} that scale invariance, definitively in a form different from (\ref{SP}), does not hold. In 1983, Parisi and Frisch \cite{frisch1985turbulence} made the following observation: the Navier-Stokes equations (\ref{NS}) are invariant under the scale transformation:
\begin{equation}
\label{scale}
 {\bf r} \rightarrow \lambda {\bf r} \, \, \, \, {\bf v} \rightarrow \lambda^h {\bf v} \, \, \, t \rightarrow \lambda^{1-h} t  \, \, \, \nu \rightarrow \lambda^{1+h} \nu
\end{equation}
Note that using (\ref{scale}) we have $\epsilon \rightarrow \lambda^{3h-1} \epsilon$.  The key observation by Parisi and Frisch is that there may exist many different value of $h$ each occurring with a probability $P(h)$, i.e. turbulent flows can be considered as a superposition of many different {\it scale} invariant configurations. In order to maintain scale invariance upon averaging over $P(h)$, one needs to assume that $P(h)  \sim r ^{F(h)}$, i.e. one needs to assume that scale invariance holds for the probability distribution of $h$. This conjecture is referred to as the {\it multifractal conjecture} because originally $F(h)$ was written in the form $3-D(h)$ where $D(h)$ is assumed to be the fractal dimension of the scale invariant solution with exponent $h$.  There exists a constrain on $D(h)$ since we must require, in agreement with (\ref{eq:anomaly}), that the energy dissipation is independent of $Re$, i.e. we must require that
\begin{equation}
\label{zeta3}
\langle \epsilon \rangle  \sim \int dr \, r^{3-D(h)} r^{3h-1} \sim const
\end{equation}
Note that eq.(\ref{SP}) is now no longer valid since we must compute the average over $h$. Because we are interested in the limit at small scale, we can use the saddle point technique to compute the integral and we obtain:
\begin{equation}
\label{MFSP}
S_p(r)  \sim \int dr \, \, r^{ph+3-D(h)} \sim r^{\zeta(p)} 
\end{equation}
where 
\begin{equation}
\label{zetap}
\zeta(p) = inf_h [ p h + 3 -D(h)]
\end{equation}
Clearly, if we know $\zeta(p)$ we can invert the Legendre transform (\ref{zetap}) to compute $D(h)$. 
At first sight the multifractal conjecture may seem rather obscure because it is not clear what is its physical meaning, i.e what does it mean  {\it averaging} over the possible scale? invariant configurations. Next, the multifractal conjecture seem not to have any predictive power because nothing is known on the function $D(h)$ except the constrain (\ref{zeta3}). Finally, the discussion on vortex filaments suggests that the function $D(h)$ may not be universal. 

\begin{figure}[h]
 \begin{center}
 \includegraphics[width=0.90\textwidth]{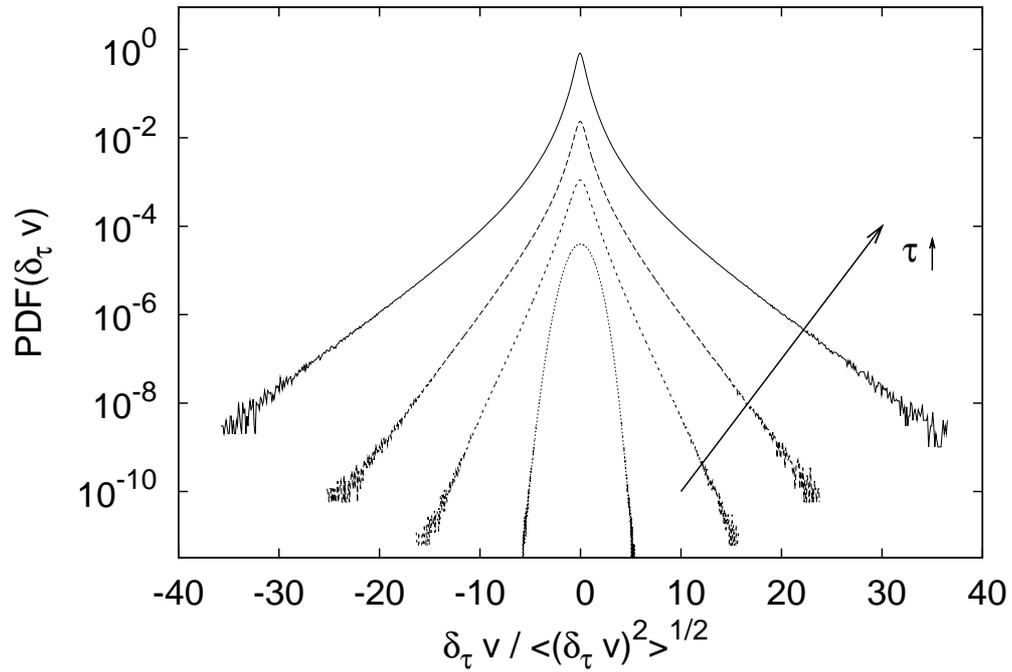}
 \end{center}
   \caption{Probability distribution functions (PDF) of the Lagrangian velocity increments at changing the time lag $\tau$. Curves have been shifted along the y-axis for the sake of presentation.For details about the numerical simulations see \cite{Benzi:2010uz}}
  \label{fig2}
\end{figure}

\section{Anomalous scaling and universality}
Using eq. (\ref{MFSP}) we can easily obtain $\Gamma_p(r) \sim r^{\zeta(p) - p\zeta(2)/2 } $. Since $\zeta(p)$ is a convex function of $p$, we obtain that $\Gamma_p(r)$ must increase for $r \rightarrow 0$, i.e. intermittency is consistent with the multifractal conjecture. The next step is to compute from experimental and numerical data $\zeta(p)$ for different forcing mechanism and $Re$ and to understand whether $\zeta(p)$ are universal. The computation of $\zeta(p)$ requires the existence of a clear  scaling range where (\ref{MFSP}) is observed. Because any numerical simulations or laboratory experiments are done for finite system sizes, we can expect that there must be non trivial effects in the scaling of  $S_p(r)$.
Such finite size effects are quite common in many systems and they are extremely well known and under control  in the case of critical phenomena. However, since we have no theory to compute $\zeta(p)$ from the Navier-Stokes equation, we are unable to predict finite size effects. This is a major problem to understand whether scaling and universality is observed for turbulent flows.

\begin{figure}[h]
 \begin{center}
 \includegraphics[width=0.50\textwidth]{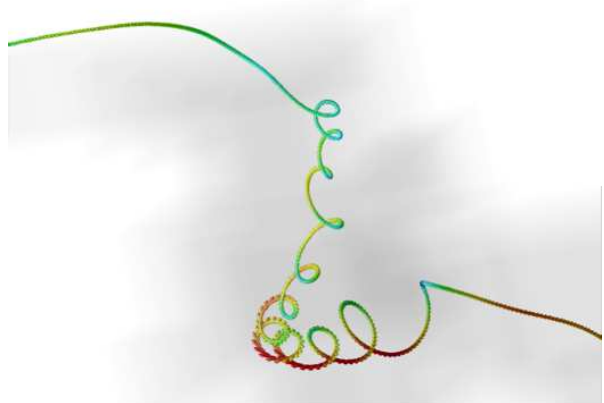}
 \end{center}
   \caption{A typical trajectory of a Lagrangian tracer in HIT inside a vortex filaments.}
  \label{fig3}
\end{figure}

\bigskip
A systematic investigation of the scaling properties of Eulerian turbulence required several years of work and the introduction of new ideas to overcome finite size scaling (see the work on Extended Self Similarity). The overall conclusion \cite{Arneodo:1996hf}  was that scaling is observed and that $\zeta(p)$ are independent on $Re$ and the forcing mechanism, see e.g.  figure (\ref{fig11}).  
This conclusion is obtained by accurate data analysis of many different laboratory and numerical data and it does not imply anything about the multifractal conjecture. In order to make progress, one needs to understand in a deeper way the physical mechanism of intermittency and its relation, if any, with the function $D(h)$.

\bigskip

A major breakthrough in this direction was provided by the exact non trivial solution of the so called {\it Kraichnan} model of a passive scalar \cite{Falkovich:2001uz}. We consider the following problem: let $\theta$ a scalar field advcted by  a velocity field ${\bf v}$ and forced by some large scale mechanism $F_{\theta}$:

\begin{equation}
\partial_t \theta + {\bf v} \cdot \nabla \theta = \chi \Delta \theta + F_{\theta}
\label{passive}
\end{equation}

\begin{figure}[h]
 \begin{center}
 \includegraphics[width=0.9\textwidth]{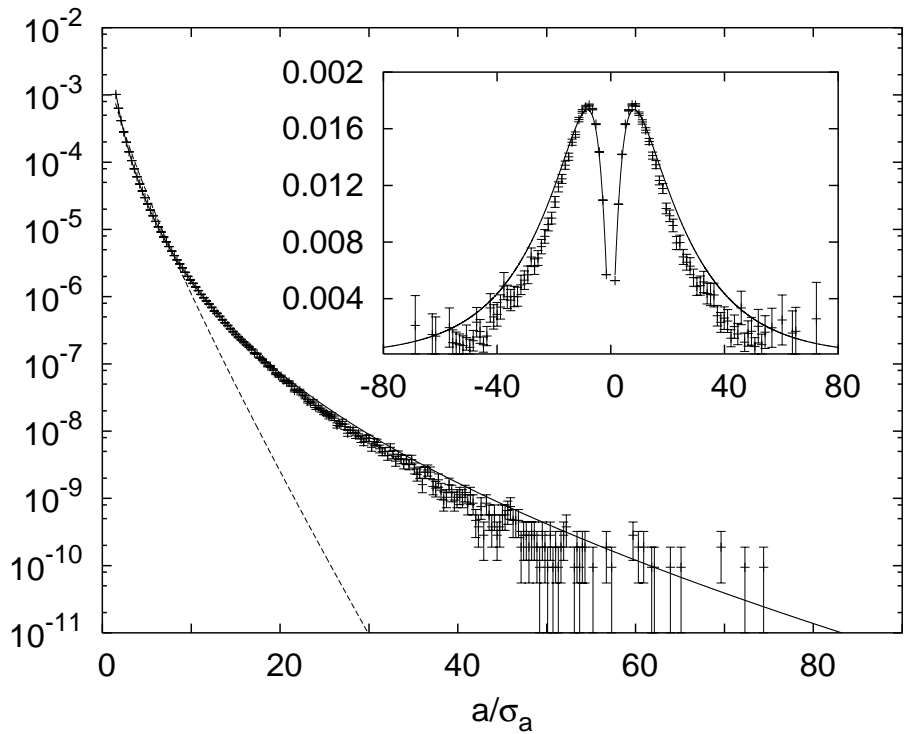}
 \end{center}
   \caption{Acceleration PDF for HIT. The dashed line represents the dimensional K41 prediction. The black continuos line is the multifratcal prediction 
\cite{Biferale:2004hj}. Inset: Acceleration PDF multiplied by fourth order ponwer of the acceleration, $a^4 P(a)$ and the corresponding prediction from the multifratcal 
formalism}
  \label{fig4}
\end{figure}

Following Kraichnan  we assume: (i) the velocity field is gaussian random field delta-correlated in time and with prescribed correlation function $g(r)  \sim r^{\xi}$; (ii) the large scale forcing is  random, gaussian, and statistically isotropic. Given (\ref{passive}) we are interested to compute the behavior of the correlation functions $C_n(x_1,x_2, ... , x_n) = \langle \theta(x_1)\theta(x_2) .. \theta(x_n) \rangle$ as  a function of $\xi$ in the limit $\chi \rightarrow 0$.  For this problem, we can rephrase the scale transformation (\ref{scale}) as:
\begin{equation}
\label{scalepassivo}
 {\bf r} \rightarrow \lambda {\bf r} \, \, \, \, {\theta} \rightarrow \lambda^{h} {\theta} \, \, \, t \rightarrow \lambda^{1-\xi} t  \, \, \, \nu \rightarrow \lambda^{1+\xi} \nu
\end{equation}
The analogous of (\ref{K45}) can be easily derived upon assuming that $\chi \langle (\nabla \theta)^2 \rangle  \sim const.$, which is equivalent to  $\xi + 2 h = 1$ (for delta-correlated random field ${\bf v}$ a subtle difference apperas that  is not important in the following discussion).  Since eq. (\ref{passive}) is linear and because the random field is delta-correlated in time, we can obtain a closed equation for $C_n$ which can be formally written as
\begin{equation}
\label{correlation}
L_n[v] C_n + D_n C_n + F_n = 0
\end{equation}
where $L[v]$ is a linear  operator depending on the velocity  statistics, $D_n$ is the n-dimensional Laplacian proportional to the diffusivity $\chi$ and $F_n$ is the term due to the forcing.  The exact form of the operators appearing in (\ref{correlation}) is irrelevant for our argument. The general soution of (\ref{correlation}) is given by $C^{(i)}_n+ C^{(d,f)}_n$ where the first term is one of the  {\it zero mode} solutions of $L[v]C^{(i)}_n=0$ while the second term is the (particular) solution depending on $\chi$ and the forcing, i.e. $C^{(i)}_n$ is the correlation function in the inertial range while $C^{(d,f)}_n$ is the non universal part of the correlation function which depends on forcing and dissipation. The important result is that in the limit $\xi \rightarrow 0$ one can show that the inertial range dynamics $C^{(i)}_n$, is the relevant contribution to the correlation function. Moreover, the $n-$order correlation function display anomalous scaling, i,e $C_n( \lambda x_1, ... \lambda x_n) \sim \lambda^{z(n)} C_n(x_1, .. x_n)$ where $z(n)$ is a non linear function of $n$. In summary, for the passive scalar problem described by (\ref{passive}), one is able to show that scaling occurs and that the scaling properties of the correlation functions are universal. 

\begin{figure}[h]
 \begin{center}
 \includegraphics[width=0.90\textwidth]{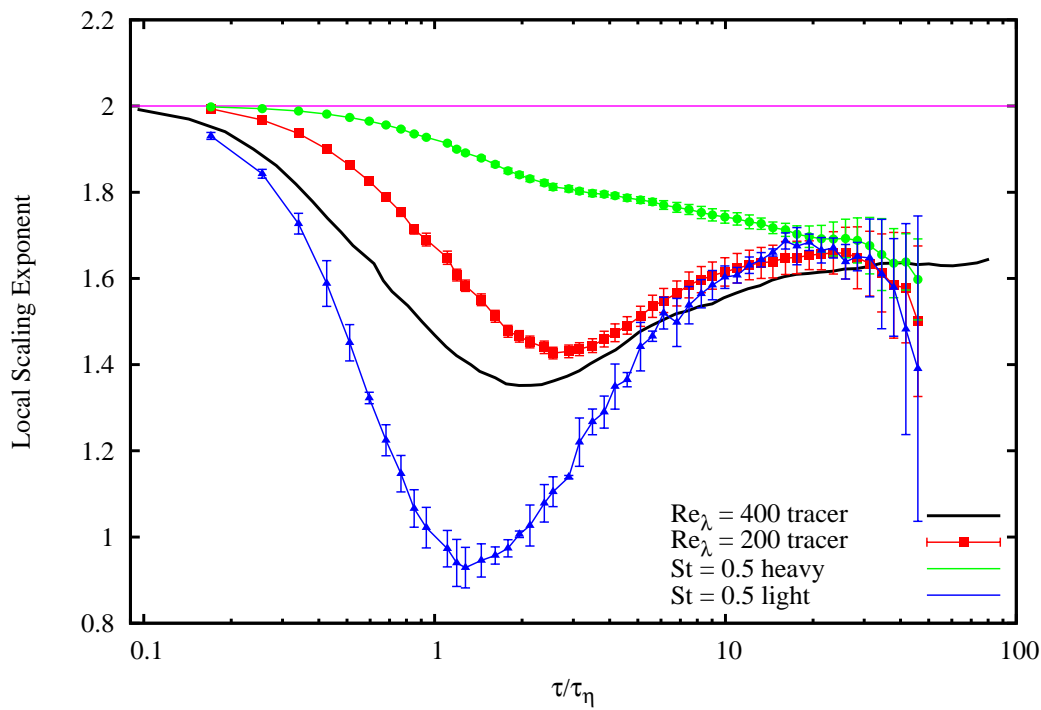}
 \end{center}
   \caption{Comparison of local scaling exponents for the 4th order Lagrangian Flatnees between: (i) tracers particles at two different Reynolds number (ii)  one light and (iii) one heavy particle. Notice the ehnancement (depletion)  of the bottleneck around $\tau/\tau_\eta \sim 1$ for light (heavy) particles with respect to the tracers' statistics. The horizontal line corresponds to the K41 non-intermittent prediction $=2$. (Figure courtesy of E. Calzavarini) }
  \label{fig8}
\end{figure}

\bigskip
One can wonder whether there exists something analogous of vortex filaments in the passive scalar. Numerical simulations show that in the case of passive scalar there are many fronts in $\theta$ (abrupt changes in $\theta$ over a very small distance) \cite{Celani:2001dz}. A front like structure corresponds to $h=0$ in the multifractal language. Therefore, if the multifractal conjecture is correct, we should expect $z(n) \rightarrow = z_{\infty} \equiv 3-D(0)$ for $n \rightarrow \infty$, where $D(0)$ is the fractal dimension of the front. This relation can be checked and it appears to be consistent with the numerical simulations. The consistency by itself is not surprising since $D(0)$ is defined as the set of point where $h=0$, i.e. where we observe a front. The interesting point is that the results on the correlation functions and the universality of $z(n)$ imply that the statistical properties of the fronts are independent on the dissipation mechanism and the large scale forcing. In other words, the statistical properties of $\theta$ can be described using the multifractal conjecture independently on the existence of well defined coherent structures (fronts) carrying the strongest singularities in  $\nabla \theta$. Moreover, the general solution of eq.(\ref{correlation}) tells us that fronts are formed due to the complex non linear interaction between the velocity field and the passive scalar {\it in the inertial range} and scale invariance is not affected by the existence of fronts. Going back to the Navier-Stokes equation we can imagine that vortex filaments are consistent with the asymptotic form of $\zeta(p) \rightarrow p h_{0} + 3-D(h_{0})$ and that the statistical properties of vortex filaments do not affect the scaling properties of the inertial range (universality). This is a delicate  statement which we now investigate in details.

\section{The dissipation range}

\begin{figure}[h]
 \begin{center}
 \includegraphics[width=0.90\textwidth]{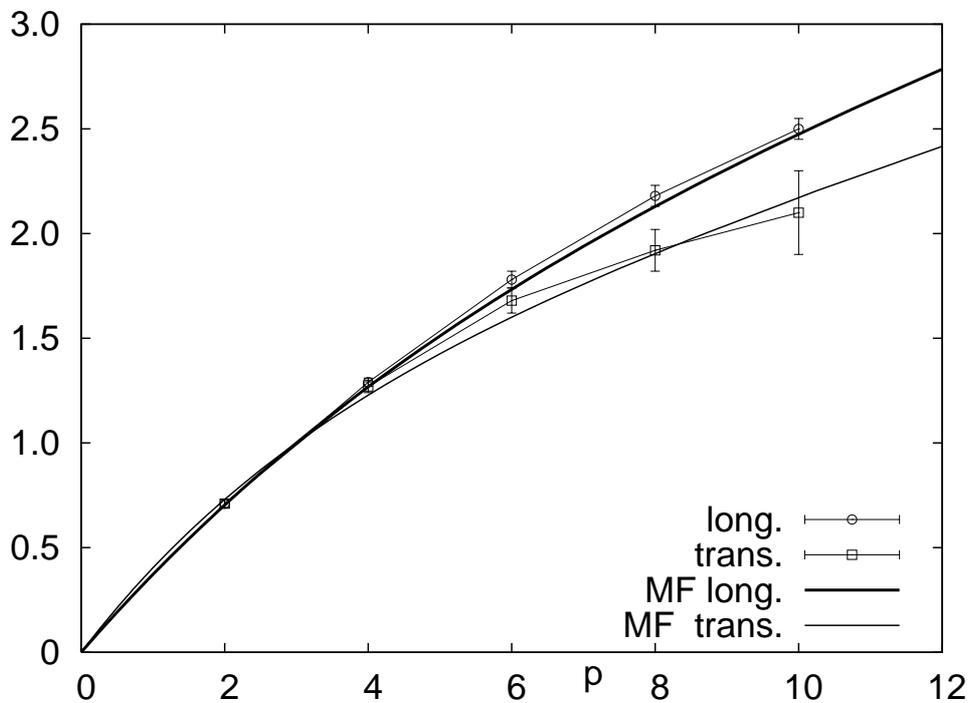}
 \end{center}
   \caption{Comparison between Eulerian scaling exponents  for longitudinal, $\zeta_l(p)$, and transverse, $\zeta_{tr}(p)$, Structure Function \cite{Benzi:2010uz} together with two different multifractal predictions (MF) obtained with two different choices of $D(h)$.}
  \label{fig10}
\end{figure}

\begin{figure}[h]
 \begin{center}
 \includegraphics[width=0.90\textwidth]{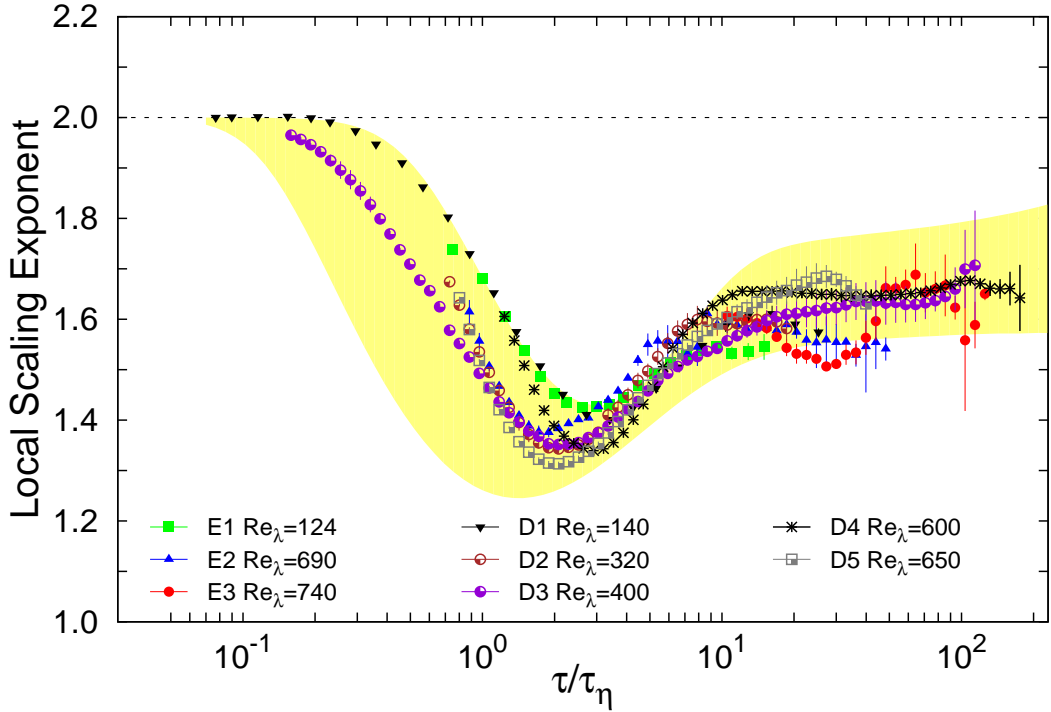}
 \end{center}
   \caption{Comparison between experimental and numerical data of local scaling exponents of Flatness (from \cite{Arneodo:2008gq}). The horizontal line corresponds to the K41 non-intermittent prediction.The yellow band is the multifractal  prediction  with uncertainty estimated out of the  two different $D(h)$ curves estimatred from either longitudinal or transverse Eulerian increments. Data are taken from \cite{Berg2006,Xu2006,Mordant2001,Yeung2009,Mordant:2004um,Homann2007,Biferale2005}}
  \label{fig11}
\end{figure}

\bigskip

Let us assume for the time being that the multifractal conjecture is correct, although we do not know how to compute $D(h)$ in the Navier-Stokes equation. We already said that any numerical and/or experimental observation shows  finite size effects. Let us focus on the dissipation effects due to the viscosity. We cannot blindly assume that the multifractal picture holds in the dissipation range ($r \ll \eta$).  We can speculate that when $\delta v(r) r / \nu \sim 1$, the velocity fluctuations are damped by viscous effects. The problem is that we must  give a reliable meaning to  $\delta v(r)$. The question is: when we write $\delta v(r) \sim r^h$ what we really mean? ( Unfortunately the solution of the passive scalar does not help in this case.) It turns out that one can give a well defined meaning to scaling $\delta v(r) \sim r^h$ using the theory of random multifractal fields. For our purpose, it is enough to say that the statistical properties of velocity fluctuations at scales $r$ and $R$ are linked by the relation (with $r < R$):
\begin{equation}
\label{multiscale}
\delta v(r) = \delta V(R) \left[ \frac{r}{R} \right]^h
\end{equation}

with propability $P_h(r/R) \sim (r/R)^{3-D(h)}$. It is possible to  construct  explicit examples of random fields obeying eq. (\ref{multiscale}) with a prescribed function $D(h)$ \cite{Benzi:1993uza}. Using (\ref{multiscale}) we can write that dissipation effects are relevant when 
\begin{equation}
\label{vergassola1}
\frac{ \delta v(r) r}{ \nu}  =  \frac{ \delta v(R) R}{ \nu}  \left[ \frac{r}{R} \right]^{1+h} \sim 1
\end{equation}
Upon choosing  $R=L$ and denoting the dissipation scale $\eta(h)$ we obtain
\begin{equation}
\label{etah}
\eta(h) =  Re^{-\frac{1}{1+h}} \, L
\end{equation}
The above relation is true with probability $(\eta(h)/L)^{3-D(h)} = Re^ {-\frac{3-D(h)}{1+h}}$.  Using (\ref{vergassola1}) and (\ref{etah}) , we can compute the average energy dissipation
$\epsilon$:
\begin{equation}
\label{consistent}
\epsilon = \int dh \nu \frac{\delta v(\eta(h))^2}{\eta(h)^2} \sim \int dh Re^{-\frac{3h+2-D(h)}{1+h}}
\end{equation}
It is easy to show that the saddle point computation of the above integral is equivalent to the condition $\zeta(3)=1$ which is a constrain on $D(h)$. Using this result, we obtain $\epsilon$ independent of $Re$, which shows the consistency of (\ref{vergassola1}) with the eq. (\ref{K45}). We can generalize eq.(\ref{consistent}) to compute the scaling behavior in $Re$ of moments of velocity gradients \cite{Nelkin:1990kz}:
\begin{equation}
\label{gradienti}
\langle (\nabla v)^p \rangle = \int dh \frac{\delta v(h)^p}{\eta(h)^p} Re^{\frac{3-D(h)}{1+h}}  \sim \left[\frac{\delta v(L)}{L} \right]^p \int dh Re^{ -\frac{p(h-1)+3-D(h)}{1+h}} \sim \left[\frac{\delta v(L)}{L} \right]^p Re^{\chi(p)}
\end{equation}
where $\chi(p) = sup_h [ -(p(h-1)+3-D(h))/(1+h)]$. From (\ref{gradienti}) it is possible to show that for $p>2$ the following inequality holds.
\begin{equation}
\label{gkurtosis}
\gamma_p \equiv \frac{\langle (\nabla v)^p \rangle}{\langle (\nabla v)^2\rangle^{p/2}} = Re^{\chi(p)-p/2} > lim_{r\rightarrow \eta} \Gamma_p(r) \sim Re^{\frac{3}{4}(\zeta(p)-p\zeta(2)/2)}
\end{equation}
The above equation tells us that  intermittency in the dissipation range $Re^{\chi(p)}$ is greater than  intermittency extrapolated from the inertial range, l.h.s of (\ref{gkurtosis}). This is a highly non trivial prediction given by multifractal conjecture and consistent with the experimental and numerical data.  

\bigskip

One can take advantage of (\ref{vergassola1}) to generalize the multifractal conjecture for finite $Re$ numbers and introducing the effect of dissipative contribution. Instead of (\ref{multiscale}) we have the following result:
\begin{eqnarray}
\label{multiv}
\delta v(r) &=& g(\frac{r}{L}, \frac{\eta(h)}{L}) f(\frac{r}{L},\frac{\eta(h)}{L})^h\\
\label{multip}
P_h(r) &=& f(\frac{r}{L},\frac{\eta(h)}{L})^{3-D(h)}
\end{eqnarray}
where the two functions $g(x,y)$ and $f(x,y)$ satisfy the following asymptotic conditions:

\begin{eqnarray}
\label{cgrande1}
lim_{y \rightarrow 0} f(x,y) &=& x,\\
\label{cgrande2}
 im_{y \rightarrow 0} g(x,y) &=& const.
\end{eqnarray}
and
\begin{eqnarray}
\label{cpiccolo1}
lim_{x \rightarrow 0} f(x,y) &=& const. \\
\label{cpiccolo2}
lim_{x \rightarrow 0}g(x,y) &=& x.
\end{eqnarray}
In the range $ \eta(h) \gg r$, we assume that no intermittent fluctuations occur and the velocity fluctuation are smooth according to (\ref{cpiccolo2}) and (\ref{multiv}), while in the inertial range $ r \gg \eta(h)$ the dissipation effects are irrelevant.  The precise shape of the function $f$ and $g$ are supposed to be universal. Presently, we are not able to compute $f$ and $g$ theoretically but we can provide very accurate fit of both functions using experimental and numerical data. At any rate, we shall see in the following that the detailed shape of the two functions do not play a crucial role in the theory. It is interesting to remark that, if we neglect the fluctuation of the dissipative scale by assuming $\eta(h)= \eta  = Re^{-3/4} L$, eq.s (\ref{multiv},\ref{multip}) predict
\begin{equation}
S_p(r) \sim g^p(r) f(r)^{\zeta(p)}
\end{equation}
In the range of scales where $g(r) \sim const$ we obtain:
\begin{equation}
\label{ESS}
S_p(r)  \sim S_3(r)^{\zeta(p)/\zeta(3)} \sim S_3(r)^{\zeta(p)}
\end{equation}
which is  known in literature as Extended Self Similarity (ESS) \cite{Benzi:1993bz,Benzi:1996wv}. ESS is useful to extract accurate values of the scaling exponents $\zeta(p)$ even at relatively low $Re$ number because it does not require any knowledge on the function $f$ and because, at low Reynolds, intermittent fluctuations of the dissipative scale are relatively small. Clearly, the range of scale where ESS is useful must be outside the dissipation range. The size of the dissipation range depends on the minimum and maximum value of $h$. Theoretically, we know that $h \in [0,1]$. Therefore we can estimate the dissipation range as $[Re^{-1}, Re^{-1/2}]$.  E.g. for $Re=10^5$ and a characteristic value of $L \sim 1 m$, the dissipation range is $[30 \mu, 3 mm]$. 

\bigskip

We can now use (\ref{vergassola1}) to predict the probability distribution of the acceleration. Let us define $\tau_{\eta}(h) = \eta(h)/\delta v_{\eta}(h)$ the dissipative time scale associated to the dissipation scale $\eta(h)$. Then, we can compute the acceleration $a(h)$ as $\delta_v{\eta}(h)/\tau_{\eta}(h)$:
\begin{equation}
\label{acc1}
a(h, \delta V(L))  = \frac{\delta V(L)^2}{L} \left[ \frac{\nu}{L \delta V(L)} \right]^{\frac{2h-1}{1+h}}
\end{equation}
which holds with probability  $(\eta(h)/L)^{3-D(h)}$. Therefore, given the probability of the large scale fluctuation $\delta V(L)$, using (\ref{acc1}) we can compute the probability $P(a)$ to observe in a given point a value $a$  for the acceleration and we can compare our findings against experimental results. In most cases, the fluctuations at very large scale $L$ are observed to be distributed in a gaussian way and we can provide an analytical  prediction for $P(a)$.  It turns out that the multifractal prediction of $P(a)$ is very accurate compared against experimental and numerical data, see figure (\ref{fig4}) and \cite{Biferale:2004hj}.  The word prediction in this case should be interpreted in the following way: given the probability distribution of large scale fluctuations {\it and  the function $D(h)$}, we can predict $P(a)$.  The function $D(h)$ can be computed from the knowledge of the scaling exponent $\zeta(p)$ which can be obtained from the available data. Therefore, we can say that from the knowledge of the intermittent fluctuations in the inertial range we can predict the probability distribution of the
lagrangian acceleration. Having saying that, the prediction of $P(a)$ is a highly non trivial result because it shows that  the multifractal framework, based on the scale invariance of the Navier-Stokes equations, correctly describes the statistical properties of turbulent fluctuations over the whole range of scales, from large scale to dissipative scales. The prediction of $P(a)$, in the sense previously discussed, represents a major achievement of our ability to provide a universal and consistent description of turbulent flows. 

\section{Vortex filaments and the multifractal conjecture}

\bigskip
As we previously discussed, strong fluctuations in the lagrangian acceleration are due to vortex filaments and are correctly described by the multifractal framework. In the multifractal approach, however, there is no point whatsoever where we introduced any physical informations concerning the existence and the relevance of coherent structures or vortex filaments. Clearly, it seems relevant to investigate this question in more details. To do so, we need to look at turbulence from the largrangian point of view. It is relatively easy to rewrite the multifractal approach in terms of lagrangian variables.  We need to consider  velocity difference between two points on the lagrangian trajectory at time interval $\tau$. The scaling property of the velocity field in terms of $\tau$ can be obtained by using the scaling relation between $r$ and $\tau$, namely $\tau  \sim r^{1-h}$ or equivalently $r \sim  \tau^{1/(1-h)}$. Then eq. (\ref{MFSP}) is generalized as follows:
\begin{equation}
\label{timeSP}
 S^L_p(\tau) \equiv \langle \delta v(\tau)^p \rangle \sim \int dh \tau^{\frac{ph}{1-h}} \, \, \tau^{\frac{3-D(h)}{1-h}} \sim \tau^{\xi(p)}
\end{equation}
Eq. (\ref{timeSP}) enables us to compute the lagrangian scaling exponents $\xi(p)$ in terms of the {\it same} multifractal function $D(h)$ used for the computation of $\zeta(p)$. In other words, we can compute the lagrangian scaling exponents by the eulerian scaling properties.  Before comparing the "prediction" given by (\ref{timeSP}) against available data, we need to discuss a subtle but non trivial question concerning isotropy. So far we have assumed that in the limit $Re \rightarrow \infty$, small scale turbulent fluctuations are isotropic. This assumption is based upon the fact that the Navier-Stokes equations are invariant under $SO(3)$ rotation group \cite{Biferale:2005cw}. However, real experimental data and/or numerical simulations are  done neither in the limit $Re \rightarrow \infty$ nor with perfect isotropic forcing. Even a small anisotropic on the large scale can introduce, at  finite $Re$, non isotropic effects at small scales. To be more quantitative, we can compute the structure functions in the Eulerian frame for longitudinal velocity difference and transverse velocity difference. Let us indicate with $\zeta_l(p)$ and $\zeta_{tr}(p)$ the corresponding scaling exponents. Isotropy implies that $\zeta_l(p) = \zeta_{tr}(p)$ is true for any $p$. Careful investigations, using high resolution numerical simulations, have shown that isotropy is verified for $p \le 6$ while at large $p$ one observes $\zeta_{tr}(p) < \zeta_l(p)$, see figure (\ref{fig10}). In principle it should possible to formulate the multifractal conjecture by introducing isotropic and non isotropic sectors. where the non isotropic contributions are subleading with respect to the isotropic ones at small scales and large $Re$. Therefore the discrepancy between $\zeta_l(p)$ and $\zeta_{tr}(p)$ is a measure of the finite size $Re$ effects. In the Eulerian farmework it is possible to disentangle isotropic contribution from the non isotropic ones. However, lagrangian structure functions are mixing both contributions and, consequently, for latge $p$ we can predict the lagragian scaling exponent $\xi(p)$ from the eulearian ones $\zeta_{l,tr}(p)$ with error bars increasing with increasing $p$. The non trivial result is that, within error bars (careful computed following the previous discussion), eq. (\ref{timeSP}) is consistent with experimental and numerical data up to $p=10$ \cite{Benzi:2010uz}.  
\bigskip

The validity of (\ref{timeSP}) allows us to investigate the dissipative effects in the lagrangian dynamics. The fundamental advantage of the lagrangian point of view can be understood by considering (\ref{vergassola1}) in the time domain: the dissipation time $\tau_d(h)$ can be defined by the relation:
\begin{equation}
\label{biferale1}
\frac{\delta v(\tau_d(h))^2 \tau_d(h)}{\nu} \sim 1
\end{equation}
which holds with probability $ \tau_d(h)^{(3-D(h))/(1-h)}$. A simple computation shows that
\begin{equation}
\label{biferale2}
\tau_d(h)  \sim Re^{ \frac{h-1}{1+h}}
\end{equation}
Eq. (\ref{biferale2}) shows that the dissipation effects in the lagrangian framework cover a range $[ Re^{-1},1]$ much larger the dissipation range in the Eulerian framework. Thus, in the lagrangian turbulence dissipation effects are magnified.  Using (\ref{biferale2}), we can generalize (\ref{multiv}) to obtain

\begin{eqnarray}
\label{multilag}
\delta v(\tau) &=& G(\frac{\tau}{T}, \frac{\tau_d(h)}{T}) F(\frac{\tau}{L},\frac{\tau_d(h)}{T})^{\frac{h}{1-h}}\\
\label{multiplag}
P_h(\tau) &=& F(\frac{\tau}{L},\frac{\tau}{\tau_d(h)})^{\frac{3-D(h)}{1-h}}
\end{eqnarray}
where the functions $G(x,y)$ and $F(x,y)$ satisfy the same asymptotic behavior of the Eulerian case.  
We now consider the local scaling exponents ruling the scale-behaviour of generalised Flatness:
\begin{equation}
\label{kp}
K_p(\tau) \equiv \frac{d log S^L_p(\tau)}{d log  S^L_2 (\tau)} 
\end{equation}
These quantities can be directly measured on data, they do not need any fitting and can be considered an estimate  of the intermittent fluctuations at changing the reference scale: for large $\tau$ (inertial range) we have $K_p(\tau) \rightarrow \xi(p)-\xi(2)/2$ while for small tau
we have $K_p(\tau) \rightarrow p$ (dissipative range),  Using $K_p(\tau)$ we can {\it quantitatively} measure the effect of vortex filaments for the intermittent fluctuations. The smart idea is to compute $K_p(\tau)$ for lagrangian particles and for inertial particles. The latters can be heavy or light particles: heavy particles are concentrated outside vortex filaments
while light particles are concentrated inside vortex filaments. It is found that $K_p(\tau)$ show a well defined deep in the dissipation range for lagrangian particles, which disappears for heavy particles and becomes deeper for light particles. Vortex filaments are clearly associated to the increase of intermittency in the dissipation range, see figure (\ref{fig8}). 

\bigskip

It is important to understand that even for $p=4$ the value of $K_p$ depends on the whole function $D(h)$ and not from the value of $\zeta(p)$ or $\xi(p)$. In other words, if we consider $K_4$, the increase of intermittency (i.e. $K_4 < \xi(4)-\xi(2)$ depends on the whole structure $D(h)$ and the shapes of the two function $G$ and $F$.  Using $K_4$ we can assess the universality of our results. Recently, a major effort was undertaken to compute $K_4$ in a number of experimental simulations and laboratory experiments. All the results, within error bars, collapse on the same universal curve. The effect of vortex filaments, if any,  is hidden in the shape of the two functions $G$ and $F$, which interpolate the
inertial range scaling $G \sim const, F \sim \tau$ and the dissipative scaling $G \sim \tau , F \sim const $. For instance the choice
\begin{eqnarray}
G = \left[ \frac{x^c}{x^c+y^c} \right]^{1/c} \\
F = \left[ x^c + y^a \right]^{1/c}
\end{eqnarray}
provides a good fit to the data with $c \sim 4$, see figure (\ref{fig11}). 

\bigskip
The above analysis tells us something extremely interesting. First of all intermettincy and scaling in small scale turbulent fluctuations are universal and  independent of the large scale mechanisms. Second, the effects of coherent structures sum up to the same statistical probability distribution for the turbulent fluctuations. Third and more important, The small scale velocity fluctuations are consistent with the scaling properties of the system. The latter can be described as the superposition of all the possible scaling exponents $h$ with the weight $r^{3-D(h)}$.  The physical meaning
of $3-D(h)$ corresponds to that of entropy in equilibrium statistical mechanics: the mutlifractal conjecture can be rephrased by saying that the statistical properties of turbulence can be obtained by summing all possible flow configurations with exponent $h$ (energy) with the number of  available configurations being $r^{3-D(h)}$. The description of turbulence in terms of coherent structures is not in contradiction with the multifractal conjecture providing  that scaling is satisfied.  The advantage of the multifractal description is that the knowledge of $D(h)$ is sufficient to derive all the statistical properties of turbulent fluctuations at all scales and times, at least for observables which are invariant respect to the same group of symmetries of the Navier-Stokes equations. 

\bigskip 

\section{Conclusions}

Turbulence is the classical {\it prototype} of a complex system: we exactly know the equation of motions but we are or were unable to describe the macroscopic behavior of the system. In the case of homogenous and isotropic turbulence, the major effort performed in the last 20 years provides us of a well defined and relatively simple picture of turbulence:  the statistical properties are scale invariant and universal characterized by strong intermittency at all scales. The only technical point left to be done is a way to compute the function $D(h)$ from the Navier-Stokes equations. We know how to compute $D(h)$ is some relatively simple although non trivial case (the passive scalar and the Burgers' equation) and there are several hopes that the computation of $D(h)$ in the Navier-Stokes equations may eventually be done following similar ideas. In summary, we believe that the computation of $D(h)$ is not linked to a new physical ideas, although it represents a challenging  problem to be solved.

\bigskip

So far we discussed the case of homogenous isotropic turbulence. However, there exist many different turbulent problems which are worthwhile to be investigated. In particular, it is interesting to consider cases where there are new physical space scale and/or time scale which appear in the system such that eq. (\ref{K45}) should be reconsidered. A non exhaustive list includes: Rayleigh Benard convection, turbulent flows with the dilute polymers, spinodal decomposition, MHD turbulence, shear flows. In some cases, the effect of non isotropic contribution should be considered and a number of new challenging questions must be answered. It is not clear whether the same arguments reviewed in this paper
can be applied to {\it all} turbulent flows. In some cases (shear turbulence) it appears that scaling argument and the multifractal conjecture are still valid. In other cases, for instance turbulent in MHD, the question is still controversial. Also, there exists the special case of turbulence in superfluids where the dissipation mechanism is definitively not captured by the standard Navier-Stokes equations. 

\bigskip
We authors thank long lasting and useful collaboration with J. Bec, G. Boffetta, E. Calzavarini, 
A. Celani, M. Cencini, A.S. Lanotte and F. Toschi.  One of us (RB) thanks the organizing committee of DFSD-2014, and in particular Francois Dubois and Stephan Fauve for the kind invitation to the conference in Paris where this paper has been presented. The work has been supported by the European Research Council under the European Community's Seventh Framework Program,  ERC Grant Agreement No 339032.







\end{document}